\documentclass{ws-procs9x6}
\newcommand{\ok}{$O(2)$ }

\begin{document}

\title{Nonequilibrium Goldstone phenomenon in Hybrid Inflation
\footnote{\uppercase{B}ased on the presentations by \uppercase{S}z.~\uppercase{B}ors\'anyi and \uppercase{D}.~\uppercase{S}exty}}

\author{Sz. Bors\'anyi, A. Patk{\'o}s, D. Sexty}

\address{Department of Atomic Physics\\
E{\"o}tv{\"o}s University, Budapest}


\maketitle

\abstracts{
We study the onset of Goldstone phenomenon in a hybrid inflation
scenario. The physically motivated range of parameters is analyzed in
order to meet the cosmological constraints. 
Classical equations of motion
are solved and the evolution through the spontaneous symmetry breaking is
followed.  We emphasize the role of topological defects that partially
maintain the disordered phase
well after the waterfall. We study the emergence of the Goldstone
excitations and their role in the onset of the radiation dominated universe.
}

\section{Introduction}

The real time dynamics of phase transitions is one of the most important topics
of nonequilibrium field theory\cite{Bodeker:2000pa} with applications
ranging from the description of relativistic heavy ion collisions to 
cosmological scenarios such as hybrid inflation\cite{Linde}. The
reheating of the universe in this framework has been addressed by several
authors\cite{Garcia-Bellido:1997wm,Copeland:2002ku,Skullerud:2002sp}.
A particularly efficient mechanism, called spinodal (tachyonic) instability 
sets a sudden end for the inflation by the so called waterfall mechanism. This
framework involves a nonthermal 
phase transition for the matter field that is coupled to the hypothetical
inflaton field.  The formation and decay of topological or domain structures
may play a crucial role in the further evolution of the
fields\cite{Copeland:2002ku,Yamaguchi:1999yp}.

In this study we concentrate on the Goldstone phenomenon as it emerges in a
hybrid inflation scenario. We consider the coupled system of the (scalar)
inflaton field ($\sigma$) and a two-component matter field ($\Phi$) in a
self-consistently expanding flat and homogeneous FRW metrics with the following
Lagrangian:
\begin{equation}
{L}=\frac12(\partial_\nu\sigma (x))^2-\frac12m_\sigma^2\sigma^2+
\frac12|\partial_\nu\Phi (x)|^2-\frac12m_\Phi^2
|\Phi |^2-\frac{\lambda}{24}|\Phi |^4-\frac12g^2\sigma^2|\Phi |^2,
\label{lagrange}
\end{equation}

We solve the classical equations of motion numerically and investigate the
dynamics of the to-be-Goldstone modes. A brief discussion is presented first
on the choice of cosmologically viable
parameters in Eq.~(\ref{lagrange}) and we recall the most important
numerical measurement algorithms. Then the onset of Goldstone phenomenon is
addressed with emphasis on the late time behaviour of the equation of state.

\section{\label{sec:parameters}Parameters}


Let us concentrate on a GUT-scale ($\sim10^{15}$ GeV) scalar field that is
reheated by the spinodal instability induced by the much lighter ($\sim10^{12}$
GeV) inflaton field. A natural choice for the GUT self coupling is $\sim1$.
The coupling to the inflaton field is chosen in the range $0.01<g<0.1$
similarly to Copeland et al\cite{Copeland:2002ku}. This physically motivated
range of parameters have to be confronted to our cosmological expectations.

There are two constraints the solution of the model should respect. First, to
ensure a successful inflation the total number of e-foldings of the scale
factor $a(t)$ should  be $\sim50\dots100$.  The second constraint stems from
the relation of the quantum fluctuations of the inflationary period to the
density fluctuations measured by the COBE experiment\cite{kolbturner}. This
relation is due to the superhorizon evolution of the fluctuations which left
the horizon during the inflation and reentered again in the recombination era:
\begin{equation}
5\times 10^{-4}=\Bigl(\frac{\delta\rho}{\rho}\Bigr)_{COBE}
=\Bigl(\frac{\delta\rho}{\rho + p}\Bigr)_{exit}.
\label{eq:cobe}
\end{equation}
($p$ is the pressure of the system).  In the terminal phase of the inflation
characterized by the Hubble constant $H$ the quantum fluctuations of the
inflaton ($\delta \sigma \sim H /2 \pi $) dominate: 
$$
\frac{\delta\rho_\sigma}{\rho_\sigma+p_\sigma}\sim\frac{\delta\sigma
V'(\sigma)}{\dot\sigma^2}=\frac{HV'(\sigma)}{2\pi\dot\sigma^2}.
$$
During the roll of the inflaton one can directly measure the size of this
combination.  The modes reentering at recombination made their exit near the
end of inflation, therefore we decided to require the COBE estimate to be
fulfilled at $\Delta N=5$ e-foldings before the inflation is terminated.

\begin{figure}[ht]
\label{fig:parametermap}
\epsfxsize=10cm
\centerline{\epsfbox{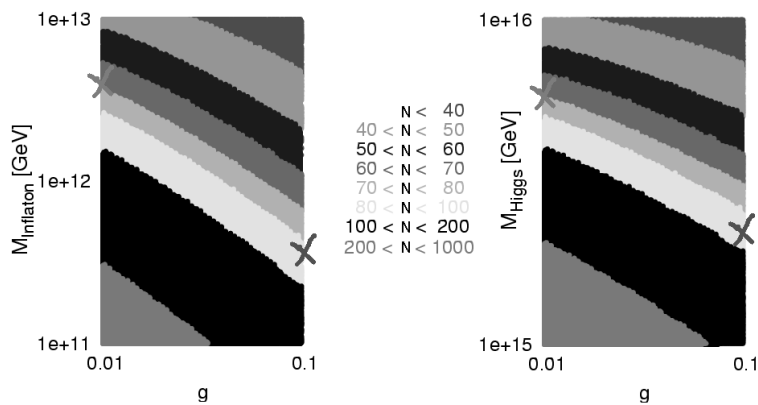}}
\caption{
Ranges of parameters that eventuate density fluctuations expected upon the
COBE results. The bands refer to different number of e-foldings ($N$). In the
present study we concentrate on the set of parameters marked by the dark cross.
}
\end{figure}

The early dynamics of the hybrid inflation is determined by the evolution of
the homogeneous mode of the inflaton field $\sigma_0(t)$ and the FRW scale
parameter $a(t)$, since the other modes are unoccupied before the onset of
spinodal instability.
At zero time the inflaton homogeneous mode is set to
$\sigma_0(t=0)=M_{\mbox{\footnotesize Planck}}$ and $\dot\sigma_0(t=0)=0$. Then
we numerically solve the Einstein equation and the
equation of motion for $\sigma_0(t)$ and $a(t)$ (the scale parameter of the
universe) without relying on the slow-roll condition. We scan through the
physically motivated range of parameters and select those models that meet the
cosmological constraints. In Fig.~\ref{fig:parametermap} we varied the
inflaton-matter coupling $g$ keeping $(\lambda/6)^{3/2}/g=1$ fixed (note that
this combination appears in the slow-roll approximation to Eq.~(\ref{eq:cobe}).

In this presentation we discuss numerical results corresponding to the
set of parameters marked by the dark cross in Fig.~\ref{fig:parametermap}.
($g=0.1$, $\lambda=1.29$, $M_H=2.2\cdot10^{15}$~GeV,
$m_{\sigma}=3.6\cdot10^{11}$~GeV.)

\section{\label{sec:methods}Measurement techniques for the physical degrees of
freedom}

\begin{figure}[ht]
\epsfxsize=10cm   
\centerline{\epsfxsize=8cm\epsfbox{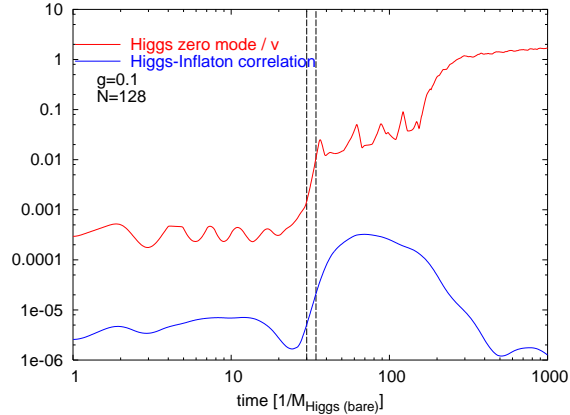}}   
\caption{Two stage symmetry breaking: in the first stage, in tachyonic 
instability, the \ok field gets excited, only later in the second
stage is the symmetry breaking completed. (The vertical lines
limit the time interval when the effective mass is negative (see text).)
The inflaton-higgs correlation squared ($W_{12}^2$) is also displayed.
}
\label{opcorr}
\end{figure}

In the \ok hybrid inflation there are three degrees of freedom at each lattice
site: the inflaton, and the two components of the \ok symmetric scalar field.
In the broken phase, our intuition is that the independent degrees of freedom
are the radial ($\rho(x)=\sqrt{\Phi_1^2(x)+\Phi_2^2(x)}$), and the angular
($\rho \phi(x)=\rho \arctan (\Phi_2(x)/\Phi_1(x)$) fields.
(One may think of the radial component as the GUT scale Higgs boson.)
In order to test the onset of this idea, we measured the velocity correlation
matrix ($W_{ij}$) which is defined by
$$ W_{ij}= \frac{\tilde{W}_{ij}}{\sqrt{\tilde{W}_{ii} 
\tilde{W}_{jj} } }, $$
\begin{equation}
\tilde{W}_{ij}=\langle v_i v_j\rangle -\langle v_i\rangle\langle v_j\rangle,
\end{equation}
where $v_i=(\dot \sigma ,\dot \rho , \rho \dot \phi)$
stands for velocity-like field components and the brackets mean 
spatial averaging. The inflaton-Higgs cross correlation can
be seen in Fig.~\ref{opcorr}. It shows a maximum after the tachyonic
instability, but it is negligible even then. Its height strongly depends on
the parameters but not on the system size.
The other two correlators (angular-Higgs and angular-inflaton) are smaller and
vanish with increasing volume.

The change of sign of the effective Higgs mass
\begin{equation}
 M^2_{\mbox{eff}}=m^2_{\Phi}+g^2 \langle\sigma^2\rangle
+\lambda\langle\rho^2\rangle/2
\label{eq:effmass}
\end{equation}
coincides with the exponential growth of the $\langle\rho\rangle$ field
(Fig.~\ref{opcorr}).  Later this effective mass relaxes close to the measured
gap value of the radial component.

The masses of the angular and radial field components were extracted from
their dispersion relation\cite{minkowskiscikk,Borsanyi:2001rc}.
The method requires uncorrelated field components in Fourier space
$X_k$. We measure the quantity  ${ |\dot X_k |^2}/{|X_k| ^2}$.
If only a single frequency ($\omega_k$) dominates the motion of a mode, than
this ratio is equal to $\omega_k^2$. Extrapolating this quantity to $k=0$ we
get the squared mass gap of the component $X$. In practice we average
over different directions in ${\bf k}$ space, over microscopical time scales
and over different runs with random initial conditions.
We measured the radial (Higgs) mass choosing 
$X=\rho=\sqrt{\Phi_1^2+\Phi_2^2}$, and the angular one with
$X=\langle\rho\rangle\exp(i\varphi)$.
We fitted the dispersion relation with a polinomial in $k^2$.
The radial mass was actually extrapolated from the $ k>0.75$ region, 
because of the observed low-$k$ region bending-down (See
Fig.~\ref{fig:disprel}).

\section{\label{sec:}Mechanism of Symmetry Breaking}


In our simulations we observe a delay in the completion of the symmetry
breaking (SB) relative to the tachyonic instability.  At the instant of the
tachyonic preheating the Higgs expectation value (vev) reaches a finite but
very small value. The symmetry is broken in the strict sense but the Higgs vev
has not reached the value one would expect from the effective potential yet. At
later times the Higgs vev approaches its final value.  (In fact, the initial
conditions break the O(2) symmetry so that we can observe the SB at the length
scale of the lattice size.  ($L\approx 90\times M_H^{-1}$)).

The sudden growth of the radial component induces large number of domains with
different directions of SB. The domain walls are characterized by small values
of the radial component.  One may also observe extended volumes of small
$\rho$, called hot spots\cite{Copeland:2002ku}.  The volume ratio of
these hot spots is reflected by the actual value of the $\rho_0(t)$ in the
period of the incomplete SB.

\begin{figure}[ht]
\epsfxsize=10cm
\centerline{\epsfbox{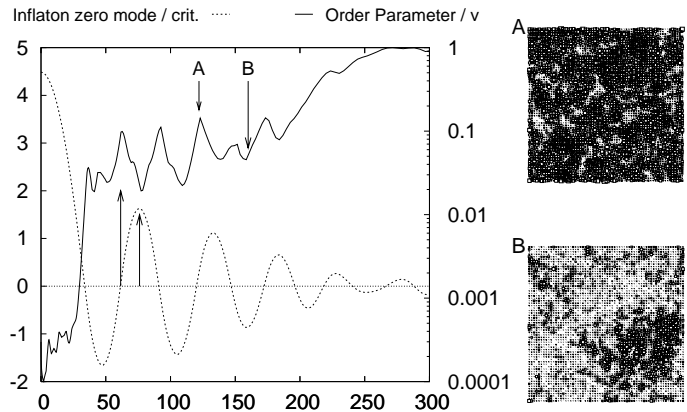}}
\caption{
Influence of inflaton zero mode oscillations (dotted, left-hand-side scale)
on the OP (solid, right-hand-side scale). At different
inflaton values lattice snapshots are displayed. The dark zones refer
to large $\rho(x)$ values. We may interpret the light zones as hot spots.
\label{fig:hotspot}
}
\end{figure}

In momentum space one can easily understand the oscillatory behavior of the OP
before the completion of the SB. The inflaton field (still oscillating) has a
direct influence on the mass squared of the matter field (see
Eq.~(\ref{eq:effmass})). The time scale of this oscillation is much slower than
the GUT scale so the elementary oscillators of the matter field are tuned
adiabatically. The amplitude squared times the frequency is an adiabatic
invariant of these oscillators.  For $\sigma_0(t)\approx0$ we have a smaller
effective mass by Eq.~(\ref{eq:effmass}) and hence, larger amplitudes and larger
value for the OP.  Opposingly, for large inflaton values the OP will come
closer to zero, this is reflected by a considerably higher volume ratio of hot
spots (see Fig.~\ref{fig:hotspot}). Note that in this regime the value of the
effective mass squared is always positive. The growth of the OP during the
completion of SB is not of tachyonic nature!

\begin{figure}[ht]
\epsfxsize=10cm
\centerline{\epsfbox{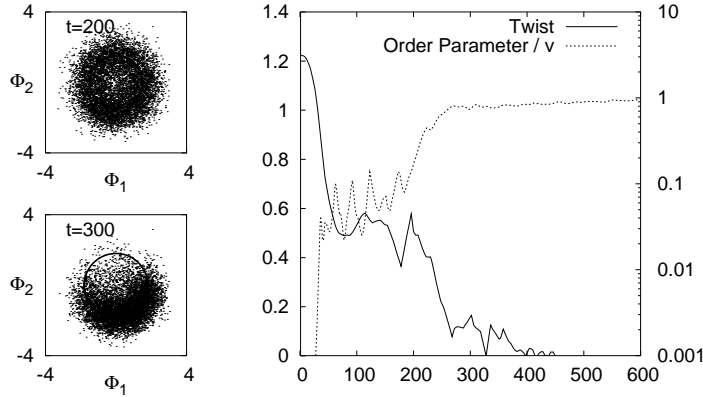}}
\caption{
The twist number
(related to the topological defect density, see text)
indicates that the decay of the domain structure coincides with the complete
breaking of symmetry. The side plots show the matter field values for the
ensemble of lattice sites right before and after the completion of SB.
\label{fig:twist}
}
\end{figure}

The OP cannot reach its broken phase value as long as the domain structure is
present. An easy way to characterize the density of topological defects is to
monitor the average of the angular twist $\Sigma_x\delta\varphi(x)$ calculated
along straight lines. This number is large right after the tachyonic
excitations of the matter field, since domains are formed with random
orientation. In the completely broken phase it should be zero. The periodic
reappearance and dissolution of hot spots erases the domain walls, and
eventually the OP departs from the vicinity of zero. As a support of this
interpretation we show the coincidence of the vanishing of the twist number and
the final growth of the OP in Fig.~\ref{fig:twist}.

After the SB is complete, the virial equilibrium of the angular and inflaton
modes is soon accomplished
($E_{\mbox{kinetic}}=E_{\mbox{gradient}}+E_{\mbox{potential}}$).
Before the onset of virial equlibrium, there is an energy excess in the angular
gradient energy, which is due to topological excitations\cite{Yamaguchi:1999yp}.

In the phase of completely broken symmetry the angular matter field components
are the Goldstone excitations. In the period of the incomplete SB, however, one
should be more careful with the identification. In the following we study how
the angular modes start obeying the Goldstone theorem.

We use now the methods reviewed in Sec.~\ref{sec:methods} to monitor the
evolution of the dispersion relation and the onset of the Goldstone theorem.

\begin{figure}[ht]
\begin{center}
\hbox{
\epsfxsize=5.5cm
\epsfbox{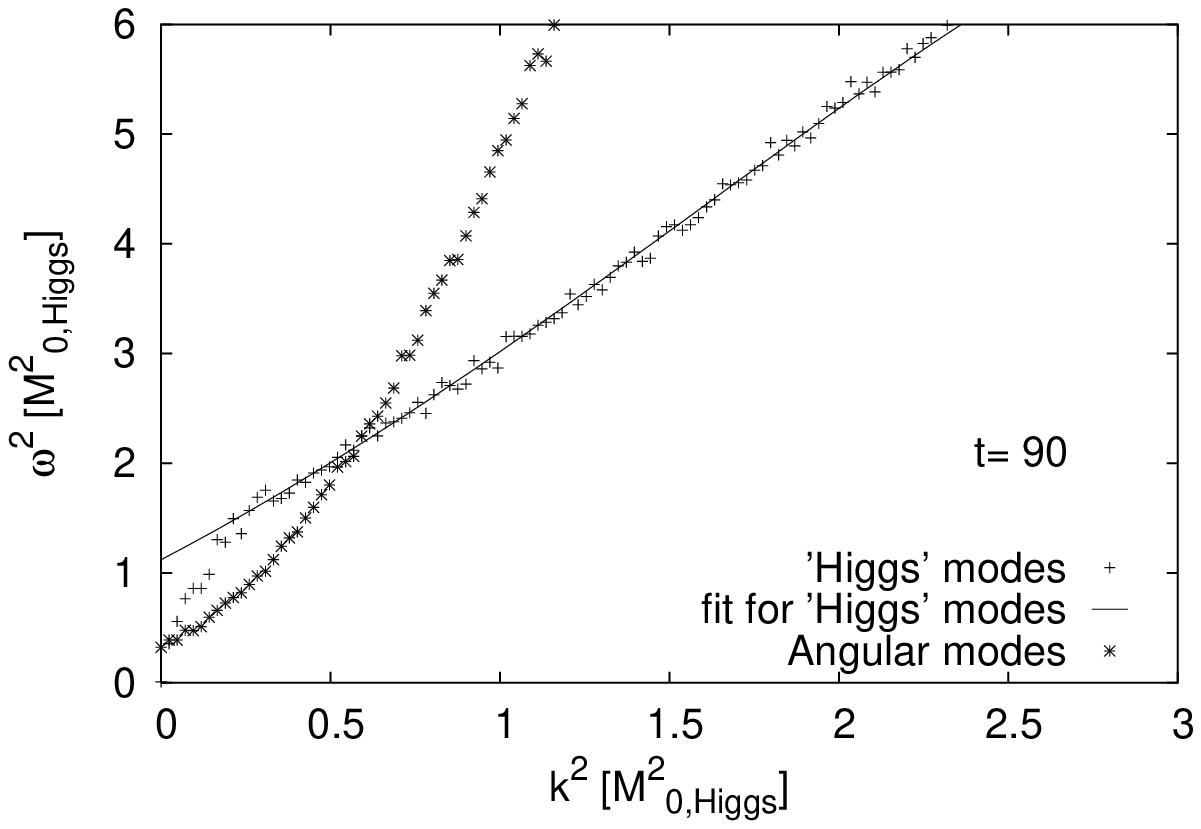}
\epsfxsize=5.5cm
\epsfbox{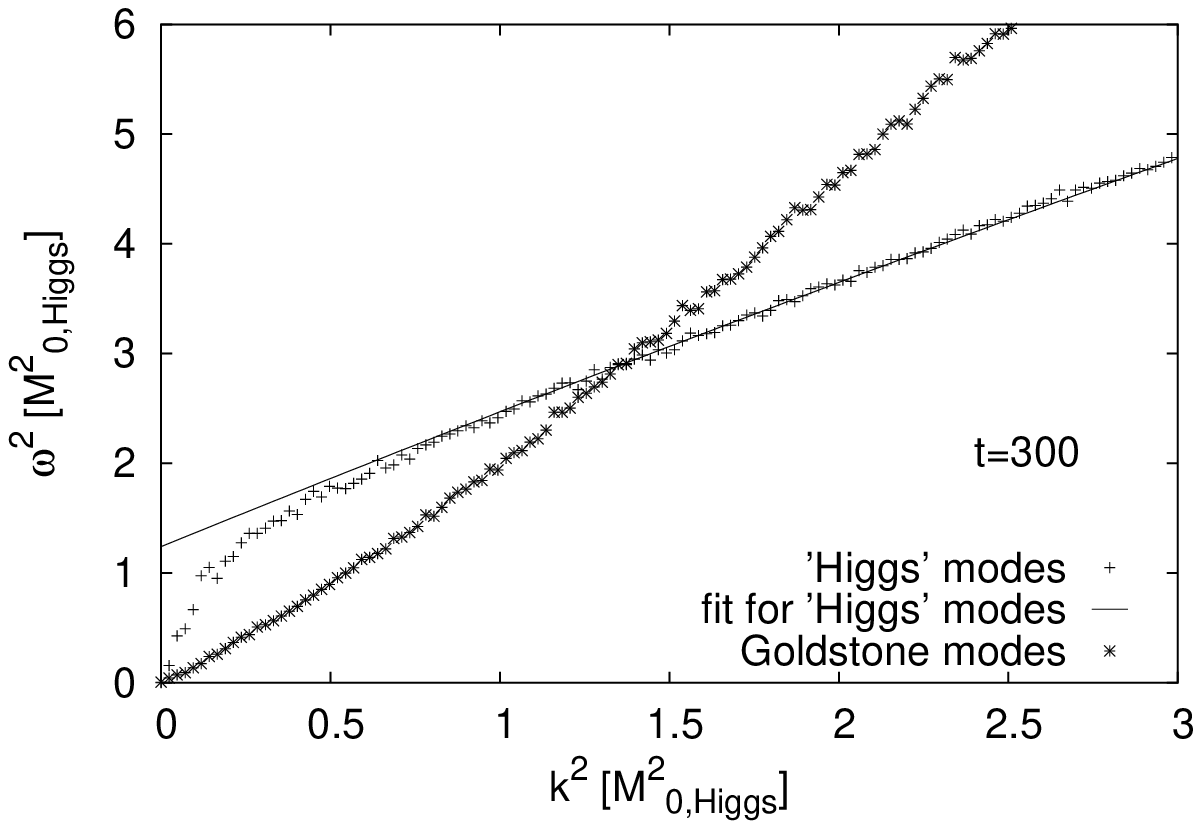}}
\end{center}
\caption{
Dispersion relations for the different matter field components before
and after the complete breaking of symmetry. On the early time plot
the angular component definitely show a nonvanishing gap, which is 
absent at late times. This fact encouraged us to refer to the angular
modes as ``Goldstone'' in the second plot.
\label{fig:disprel}
}
\end{figure}

In Fig.~\ref{fig:disprel} we display the dispersion relation for the
radial and angular field components in the partially and completely broken
phase. These curves would be straight lines for the physical quasiparticle
excitations of the equilibrium theory.  
The curvature of the dispersion relation mostly indicate the
far-from-equilibrium field configuration.
The method obviously fails with the radial (Higgs) mode for $k^2<0.5$

\begin{figure}[ht]
\epsfxsize=8cm
\centerline{\epsfbox{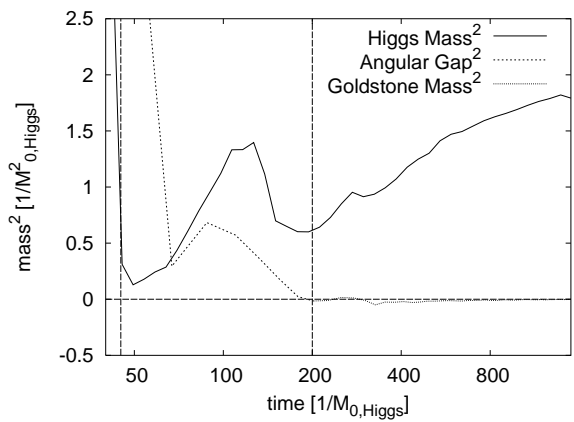}}
\caption{
Evolution of the mass gap for the radial (Higgs) and angular field components.
The decay of the domain structure coincides with the onset of the Goldstone
theorem. 
\label{fig:gaps}
}
\end{figure}

Fig.~\ref{fig:disprel} (left) clearly demonstrates a nonvanishing gap at early
times.  This gap, however, is absent in the dispersion relation measured in the
broken phase. The gap for both field components may be displayed as a function
of time if the measurements are carried out frequently through the evolution.
The resulting plot is shown in Fig.~\ref{fig:gaps}. 

As soon as the OP departs from zero (see Fig.~\ref{fig:hotspot})
the mass of the angular mode vanishes and stays further in 
the vicinity of zero.  In the presence of domains the mass scale is set by the
domain size\cite{Borsanyi:2001rc}. To have a gapless excitation this size
should grow beyond the lattice size. 

\section{\label{sec:latetime}Equation of state}

Equilibrium fields of definite mass obey well defined equations of state (EOS):
ultrarelativistic particles obey $\rho=3p$,
while for the nonrelativistic matter $p=0$. (Here $\rho$
stands for the energy density, $p$ for the pressure.) 
The appeareance of well-defined EOS gives a hint how the system arrives to the
state of the Hot Universe from the inflaton.

\begin{figure}[t]
\centerline{
\epsfxsize=5.5cm
\epsfbox{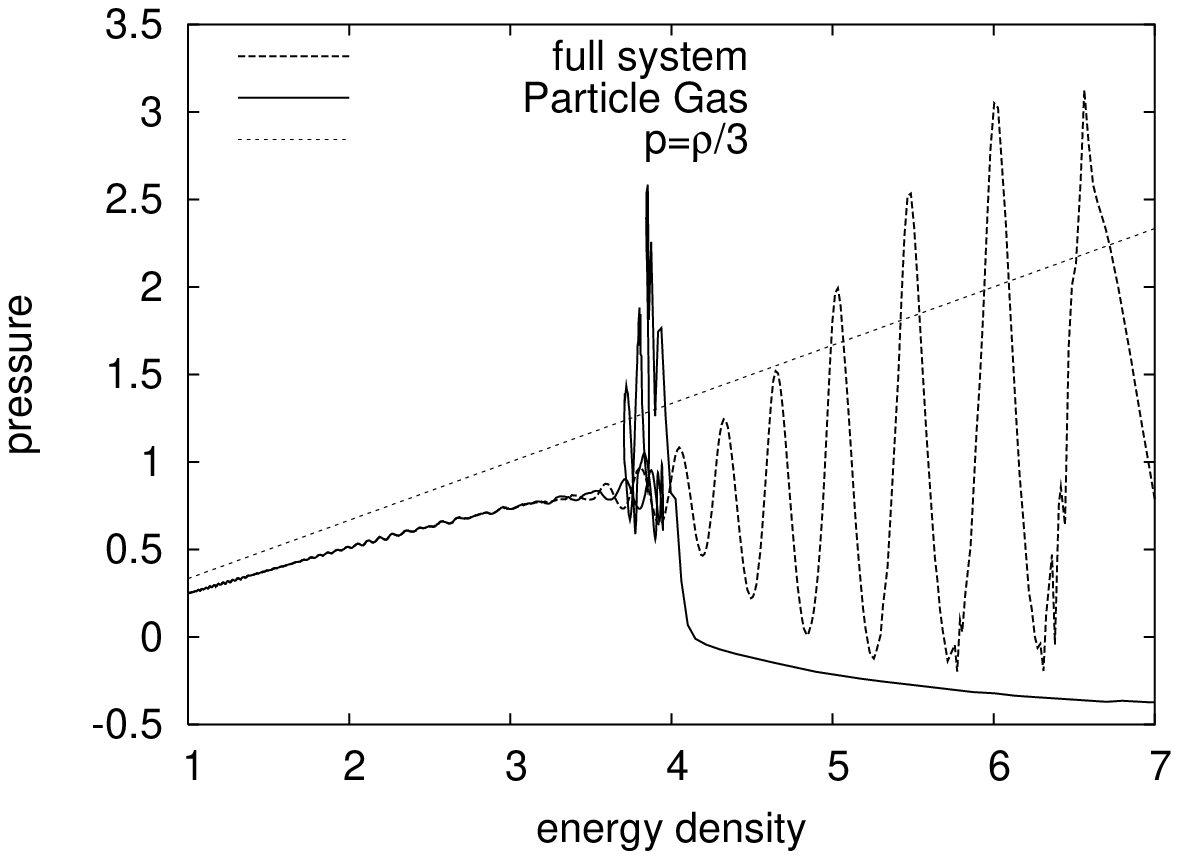}
\epsfxsize=5.5cm
\epsfbox{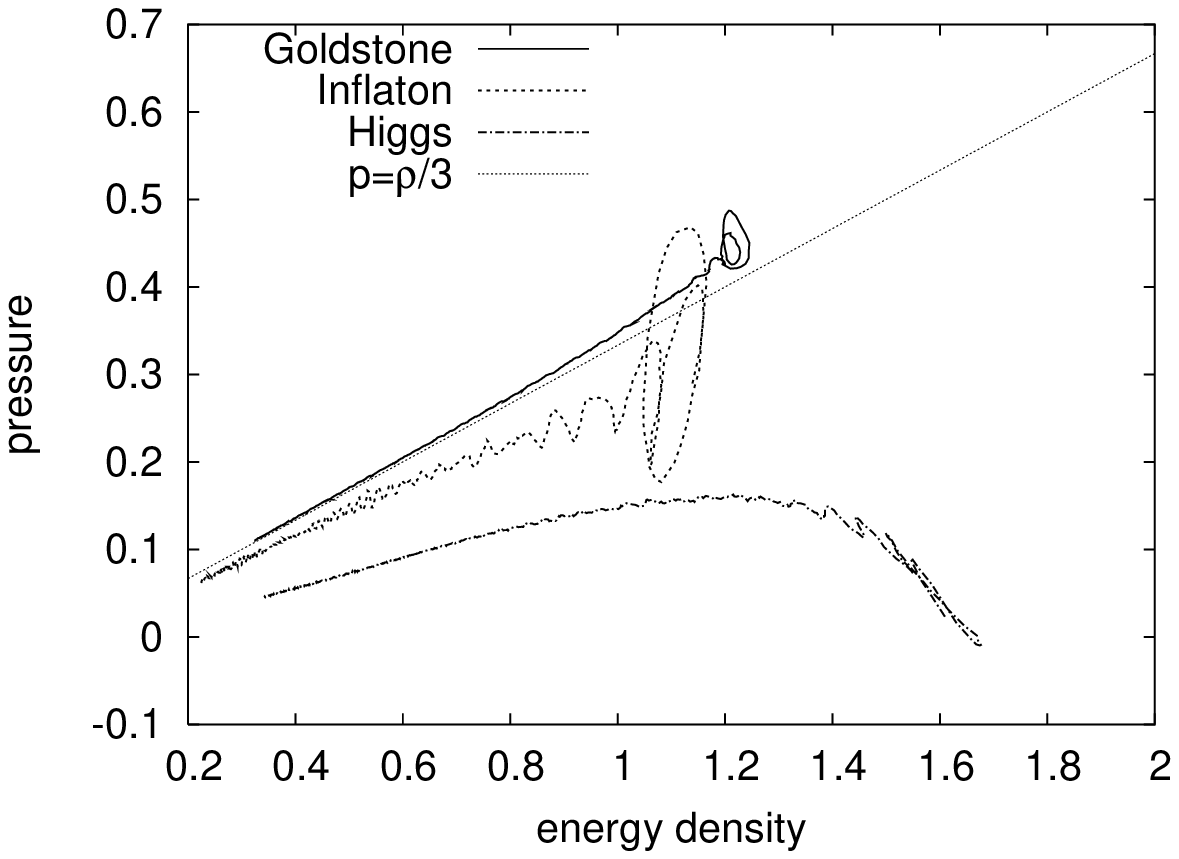}}
\caption{
Effective equations of state for the full system and the different field
compontents (see text).
\label{fig:eos}
}
\end{figure}

In a nonequilibrium evolution the EOS varies strongly with time. 
In Fig.~\ref{fig:eos} (left) we displayed the effective EOS for the full
system, which is wildly oscillating. These oscillations are completely averaged
out by the solution of the Einstein equations.

The early evolution of the energy density is dominated by the inflaton zero
mode. If one is interested in the bulk thermodynamics of the microscopic
degrees of freedom, this oscillating contribution has to be removed.  The
contribution of the inflaton zero mode has to be substracted with great care
since the potential energy for the matter field is mostly determined by the
inflaton zero mode. By making use of the separation of time scales the
oscillating part of the inflaton energy is considered to be a contribution of
the zero mode, the rest is regarded as the energy contribution of the
microscopic degrees of freedom. The equation of state for the disordered motion
of the system is also shown in Fig.~\ref{fig:eos} (left).  In
Fig.~\ref{fig:eos} (right) we show the EOS for the different microscopic field
components. The curves give the $p-\rho$ paths the field components move along,
they are displayed starting shortly before the completion of the SB.

A massive field is expected to fulfil $\rho>3p$, i.e. the corresponding curve
should be below the marked straight line in Fig.~\ref{fig:eos} (right).  The
curve for the angular component lies slightly above this line. Note that
extrapolating the dispersion relation from the high-$k$ modes one would get a
negative squared effective mass (see Fig.~\ref{fig:disprel}).  It is not the
gap itself but the entire dispersion relation that determines the EOS the
system follows.

Remarkably, the effective EOS for the light fields rapidly converge to the
expected straight line behaviour  even though the spectral equilibrium is far
from being reached.  To that time the virial equilibrium is estabilished,
although the mode temperatures are different.
In fact, in a weakly coupled massless theory each of the elementary oscillators
provide the same contribution to $\rho/p$ even if they would assume different
mode temperature\cite{Yamaguchi:1999yp}.  In conclusion we can say that the
Goldstone excitations begin to play their cosmological role right after the
complete SB by yielding a dominant contribution to the pressure and therefore
leading the universe into the radiation dominated epoch.

%
%
%
%


\begin{thebibliography}{99}

\bibitem{Bodeker:2000pa}
D.~Bodeker,
Nucl.\ Phys.\ Proc.\ Suppl.\  {\bf 94} (2001) 61


\bibitem{Linde}
A.~D.~Linde,
``Lectures on inflationary cosmology,''
arXiv:hep-th/9410082;
A.~D.~Linde,
Phys.\ Rev.\ D {\bf 49} (1994) 748

\bibitem{Garcia-Bellido:1997wm}
J.~Garcia-Bellido and A.~D.~Linde,
Phys.\ Rev.\ D {\bf 57} (1998) 6075

\bibitem{Copeland:2002ku}
E.~J.~Copeland, S.~Pascoli and A.~Rajantie,
Phys.\ Rev.\ D {\bf 65} (2002) 103517

\bibitem{Skullerud:2002sp}
J.~Skullerud, J.~Smit and A.~Tranberg,
\textit{W particle distribution in electroweak tachyonic pre-heating},
(contribution to this Workshop),
arXiv:hep-ph/0210349.

\bibitem{Yamaguchi:1999yp}
M.~Yamaguchi,
Phys.\ Rev.\ D {\bf 60} (1999) 103511

\bibitem{kolbturner}
E.~W.~Kolb and M.~S.~Turner,
``The Early Universe,'' Addison-Wesley Publishing Company, 1990

\bibitem{minkowskiscikk} Sz.~Bors\'anyi, A.~Patk{\'o}s, D.~Sexty
Phys. Rev. {\bf D 66} (2002) 025014

\bibitem{Borsanyi:2001rc}
Sz.~Bors\'anyi, A.~Patk{\'o}s, D.~Sexty and Zs.~Sz{\'e}p,
Phys.\ Rev.\ D {\bf 64} (2001) 125011

\bibitem{Felder:2000hj}
G.~N.~Felder, J.~Garcia-Bellido, P.~B.~Greene, L.~Kofman, A.~D.~Linde and I.~Tkachev,
Phys.\ Rev.\ Lett.\  {\bf 87} (2001) 011601


\end{thebibliography}
\end{document}